\documentclass[useAMS,usenatbib,onecolumn,usegraphicx]{mn2e}

\title[Softening and Reheating of the Cosmic Plasma]
  {Recombination Induced Softening and Reheating of the Cosmic Plasma}
\author[P.~K.~Leung, C.~W.~Chan, M.~-C.~Chu]
  {P.~K.~Leung$^{1,2}$\thanks{E-mail: poleung@astro.uiuc.edu}, 
  C.~W.~Chan$^{1,3}$\thanks{E-mail: cwc243@nyu.edu}, 
  M.~-C.~Chu$^1$\thanks{E-mail: mcchu@phy.cuhk.edu.hk} \\
  $^1$Physics Department, The Chinese University of Hong Kong, Hong Kong \\
  $^2$Department of Astronomy, University of Illinois at Urbana-Champaign, 
      Urbana, IL 61801, USA \\
  $^3$Department of Physics, New York University, NY 10003, USA}

\begin{document}

\date{Accepted XXXX XXX XX. Received XXXX XXX XX}

\pagerange{\pageref{firstpage}--\pageref{lastpage}} \pubyear{2003}

\maketitle

\label{firstpage}

\begin{abstract}
The atomic recombination process leads to a softening of the matter 
equation of state as reflected by a reduced generalized adiabatic
index, with accompanying heat release.  We study the effects of 
this recombination softening and reheating of the cosmic plasma on 
the ionization history, visibility function, Cold Dark Matter (CDM) 
transfer function, and the Cosmic Microwave Background (CMB) spectra.  
The resulting modification of the CMB spectrum is 1/10 of {\it WMAP}'s 
current error and is comparable to {\it PLANCK}'s error. Therefore, 
this effect should be considered when data with higher accuracy are analysed.
\end{abstract}


\section{Introduction}
Recent advances in the measurement of Cosmic Microwave Background 
(CMB) Anisotropies have attracted considerable interest in the 
field (for latest CMB observations, please refer to \citet{wmapResult}, 
website of {\it WMAP} Satellite\footnote{Website of {\it WMAP} is 
http://map.gsfc.nasa.gov/}, website of {\it PLANCK} Satellite%
\footnote{Website of {\it PLANCK} is 
http://astro.estec.esa.nl/SA-general/Projects/Planck/}, 
Max Tegmark's CMB experiment website%
\footnote{Max Tegmark's CMB experiment website is 
http://www.hep.upenn.edu/$\sim$max/cmb/experiments.html}; 
for recent reviews of CMB theory, please read 
\citet{araa} and \cite{durrer}). The cosmic 
microwave photons are believed to be remnants of radiations produced 
in the hot early universe and carry information about the state of 
the universe during the recombination epoch, when the photon cross 
section is drastically reduced due to the formation of neutral atoms. 
It is therefore possible to put tight constraints on cosmological 
theories using high precision CMB Anisotropies data, provided that 
the theoretical computation of the CMB Anisotropies also achieves at 
least the same level of precision. We need to understand the 
recombination process well if we want to extract information from 
CMB Anisotropies correctly.

In the standard calculation of the recombination process 
\citep{SSS99, SSS00}, the baryons are treated as ideal gas with an 
adiabatic index of $\gamma = 5/3$. However, as the protons and 
electrons combine to form neutral atoms, the equation of state (EOS) 
of matter is softened, as reflected by a reduced generalized 
adiabatic index \citep{chandrasekhar39,mihalas}, which depends on 
the ionization fraction. This change in the generalized adiabatic 
index is most drastic during the recombination, and the heat produced 
delays the completion of the process. In this paper, we study the 
effect of this recombination softening and reheating of the cosmic 
plasma on the ionization history, visibility function, the CMB 
spectra, and the cold dark matter (CDM) power spectra.

We first present a brief review of the standard recombination calculation
in Section~\ref{sec:standard}, which is then followed by a discussion of 
the modifications of the standard equations to take into account the 
recombination softening and reheating in Section~\ref{sec:modification}.
We then present our numerical results in Section~\ref{sec:results}, and 
summarize our discussion in Section~\ref{sec:summary}.

\section{Standard recombination calculation}\label{sec:standard}

\subsection{Expansion of the universe}

The recombination process did not happen instantaneously because of
finite interaction and transition time. Thus we need to follow
the time evolution of the ionization fraction of hydrogen and helium and
also the baryon temperature. The rate equations can be written in
terms of time $t$, which is related to the redshift $z$ by
\begin{equation}\label{eq:redshiftTime}
  \frac{{\rm d} z}{{\rm d}t}=-(1+z)H(z) \; ,
\end{equation}
where $H(z)$ is the Hubble parameter. The expansion of the universe as
a whole is described by the cosmic scale factor $a(t) = 1/(1+z)$.

The evolution of the Hubble parameter is determined by the Friedmann
equation
\begin{equation}\label{eq:friedmann}
  H(z)^2=H_0^2\left[\Omega_{\rm R}(1+z)^4
               +\Omega_{\rm M}(1+z)^3
               +\Omega_{\rm K}(1+z)^2
               +\Omega_\Lambda(1+z)^{3(1+w)}\right] \; ,
\end{equation}
where $H_0$ is the present value of the Hubble parameter and can be 
written as $100h {\rm \, km \, sec^{-1} \, Mpc^{-1}}$. The $\Omega$'s 
are the density parameters, with subscripts ${\rm R}$ referring to the 
radiation contribution, ${\rm M}$ the matter contribution (including both 
baryon ${\rm B}$ and cold dark matter ${\rm C}$), ${\rm K}$ the 
curvature contribution, and $\Lambda$ the dark energy contribution. 
$w$ is the EOS index of dark energy, which is assumed to be constant. 
The radiation density parameter is calculated by
\begin{equation}\label{eq:omegaR}
  \Omega_{\rm R}=\frac{8 \pi G (1+f_\nu) U}{3 (H_0 c)^2} \; ,
\end{equation}
where $f_\nu = 3\times(7/8)\times(4/11)^{4/3} \approx 0.681$ is the
contribution of neutrinos to the energy density of photons assuming
three massless neutrino types, $G$ is the gravitational constant, and
$U$ is the photon energy density given by Stefan's law $U = a_{\rm R}
T_{\rm R}^4$.

\subsection{Ionization fractions}

We now consider the equations for the ionization fractions. Due to the
higher ionization energy of helium comparing to hydrogen, helium 
recombined first. A helium atom has two electrons, and so there are two
processes of recombination, namely \mbox{He\,{\sc ii}} recombination
(from ${\rm He^{2+}}$ to ${\rm He^+}$) and \mbox{He\,{\sc i}}
recombination (from ${\rm He^+}$ to He atom). This \mbox{He\,{\sc ii}} 
recombination process is so fast that it can be approximated by using 
the Saha equation \citep{SSS99, SSS00}, at least in high-$\Omega_{\rm B}$ 
models. For low-$\Omega_{\rm B}$ models, the \mbox{He\,{\sc ii}} 
recombination is slightly slower than the Saha recombination. But 
since the \mbox{He\,{\sc ii}} recombination occured too early and 
had negligible effect on the CMB power spectrum, it should be 
adequate to calculate \mbox{He\,{\sc ii}} recombination by using the 
Saha equation.

The Saha equation for \mbox{He\,{\sc ii}} recombination
is\footnote{This form of the Saha equation is different from the usual
one (e.g.~in \citet{ma95}). Here we have followed the notation of
\citet{SSS99} and \citet{SSS00}. The left hand side is different
because the ionization fractions are all defined relative to
hydrogen.}
\begin{equation}
  \frac{(x_e-1-f_{\rm He})x_e}{1+2f_{\rm He}-x_e}
  = \frac{(2 \pi m_e k T_{\rm M})^{3/2}}{h^3 n_{\rm H}}
    {\rm e}^{-\chi_{\rm HeII}/k T_{\rm M}} \; ,
\end{equation}
where $x_e = n_e/n_{\rm H}$ is the ionization fraction with $n_e$
being the electron number density, $n_{\rm H}$ the total number
density of hydrogen nuclei, $f_{\rm He} \equiv n_{\rm He}/n_{\rm H} =
Y_{\rm p}/4(1-Y_{\rm p})$ the ratio of the numbers of helium nuclei to 
hydrogen nuclei, $n_{\rm He}$ the total number density of helium nuclei, 
and $Y_{\rm p}$ is the primordial He abundance by mass. Here, $m_e$ is the mass
of electron, $k$ is Boltzmann's constant, $T_{\rm M}$ is the matter
temperature\footnote{Matter is coupled by S-wave scattering
\citep{steen}. By comparing the scattering rate and the Hubble
expansion rate, it can be proved that the assumption of a single
matter temperature is valid at least down to a redshift of several
hundreds.}, $h$ is Planck's constant, and $\chi_{\rm HeII}$ is the
ionization potential of ${\rm He^{2+}}$.

For H and \mbox{He\,{\sc i}} recombinations, the rate equations 
should be used. A common method for the calculation of the 
recombination process is sometimes called the Peebles recombination 
\citep{peebles68, peebles93}. This method considers a three-level atom
with the ground state, first excited state and continuum. A
recombination coefficient $\alpha$ is used in the calculation. If one
excludes recombinations to the ground state and assumes the Lyman
lines to be optically thick, the recombination is known as Case B
recombination \citep{HSSW95,SSS00}.

Here we consider hydrogen first. The treatment for helium is
essentially the same. The ionization rate is given by
\begin{equation}\label{eq:dHdt}
  \frac{{\rm d}x_p}{{\rm d}t} 
  = C_{\rm r} \left[ \beta_{\rm H}(T_{\rm M}) (1-x_p) 
      {\rm e}^{-h \nu_{{\rm H} \,2s}/k T_{\rm M}}
    - n_{\rm H} \alpha_{\rm H}(T_{\rm M}) x_e x_p \right]
\end{equation}
\citep{peebles68,peebles93,ma95,SSS99,SSS00}. Here $C_{\rm r}$ is a
reduction factor (to be discussed below), 
$\nu_{{\rm H} \, 2s}$ is the ${\rm Ly\alpha}$ frequency, $\alpha_{\rm
H}(T_{\rm M})$ is the Case B recombination rate for hydrogen, and the total
photoionization rate $\beta_{\rm H}(T_{\rm M})$ is given by
\begin{equation}\label{eq:beta}
  \beta_{\rm H}=\alpha_{\rm H} 
                \left(\frac{2 \pi m_e k T_{\rm M}}{h^2}\right)^{3/2} 
                {\rm e}^{-h \nu_{{\rm H} \,2s}/k T_{\rm M}} \; .
\end{equation}

Due to the large Lyman alpha and Lyman continuum opacities,
recombination directly to the ground state will lead to immediate
re-ionizaton \citep{peebles68, ma95}. We therefore do not include
recombination directly to the ground state.  Recombination occurs
either by 2s level to ground state decay (with rate $\Lambda_{\rm
H}$), or by cosmological redshifting of Lyman alpha photons (with rate
$\Lambda_\alpha$). Since an atom in the 2s level may be ionized before
decaying to ground state, the recombination rate is reduced. The
reduction factor $C_{\rm r}$ is the ratio of the net decay rate to the 
sum of the decay and ionizaton rates from the 2s state, given by
\begin{equation}\label{eq:CrFactor}
  C_{\rm r}=\frac{\Lambda_\alpha + \Lambda_{\rm H}}
           {\Lambda_\alpha + \Lambda_{\rm H} + \beta_{\rm H}(T_{\rm M})} \; ,
\end{equation}
where
\begin{equation}
  \Lambda_\alpha=\frac{8 \pi}{a \lambda_\alpha^3 n_{1s}} 
    \frac{{\rm d}a}{{\rm d}t}\; , \;
  \lambda_\alpha=\frac{8 \pi \hbar c}{3 \times 13.6~{\rm eV}} \; .
\end{equation}
It is a very good approximation to replace $n_{1s}$ by 
$(1-x_{\rm H})n_{\rm H}$ for $T_{\rm M} \ll 10^5~{\rm K}$.

Now all we need are the recombination coefficients for H and
\mbox{He\,{\sc ii}}. The recombination coefficient of hydrogen 
$\alpha_{\rm H}$ is given by the fitting function \citep{SSS00}
\begin{equation}\label{eq:HCoeff}
  \alpha_{\rm H}=10^{-19} 
                 \frac{a t^b}{1+ct^d}
                 {\rm \, m^3 \, s^{-1}} \; ,
\end{equation}
with $a=4.309$, $b=-0.6166$, $c=0.6703$, $d=0.5300$, and 
$t=T_{\rm M}/10^4~{\rm K}$. The recombination coefficient of 
\mbox{He\,{\sc i}} $\alpha_{\rm HeI}$ is fitted as\footnote{There is
a typo error in \citet{SSS99} Eq.~4.}
\begin{equation}\label{eq:HeCoeff}
  \alpha_{\rm HeI}=q \left[\sqrt{\frac{T_{\rm M}}{T_2}}
                   \left( 1+\sqrt{\frac{T_{\rm M}}{T_2}} \right)^{1-p}
                   \left( 1+\sqrt{\frac{T_{\rm M}}{T_1}} \right)^{1+p}
                   \right]^{-1}
                   {\rm \, m^3 \, s^{-1}} \; ,
\end{equation}
with $q=10^{-16.744}$, $p=0.711$, $T_1=10^{5.114}~{\rm K}$ and $T_2$
fixed arbitrarily at 3 K.

Note that in \citet{SSS99}, a fudge factor 1.14 is multiplied to
Eq.~\ref{eq:HCoeff} to approximate the result of a more detailed
calculation. The detailed calculation includes totally 609
levels. There are 300 levels of H, 200 for \mbox{He\,{\sc i}}, 100 for
\mbox{He\,{\sc ii}}, 1 each for \mbox{He\,{\sc iii}},  electrons, 
protons, and each of the five molecular or ionic H species. The
purpose of our work is not to reproduce the 609-level result. Since we
just want to investigate the effect of taking into account the
ionization energy of particles, and using the fudge factor will give
unnecessary complication of the problem, we set the fudge factor to
1. Then the whole set of equations reduces to the Peebles
recombination.

We can now write down the two ionization fraction rate equations for
the proton fraction $x_p$ and the singly ionized helium fraction
$x_{\rm HeII} = n_{\rm HeII}/n_{\rm H}$, as described by
\citep{SSS99}
\begin{equation}\label{eq:HRate}
  \frac{{\rm d}x_p}{{\rm d}z} 
  = \frac{\left[ x_e x_p n_{\rm H} \alpha_{\rm H}
          - \beta_{\rm H} (1-x_p) 
          {\rm e}^{-h \nu_{{\rm H} \,2s}/k T_{\rm M}} \right]
          \left[ 1 + K_{\rm H} \Lambda_{\rm H} n_{\rm H} 
          (1-x_p) \right]}
         {H(z) (1+z) \left[ 1 + K_{\rm H} (\Lambda_{\rm H}
          + \beta_{\rm H})
          n_{\rm H}(1-x_p) \right] } \; ,
\end{equation}
and
\begin{equation}\label{eq:HeRate}
  \frac{dx_{\rm HeII}}{dz}
     = 
  \frac{\left[ x_{\rm HeII} x_e n_{\rm H} \alpha_{\rm HeI}
        - \beta_{\rm HeI} (f_{\rm He} - x_{\rm HeII})
        {\rm e}^{-h \nu_{{\rm HeI\,2^1}s}/k T_{\rm M}} \right]
        \left[ 1 + K_{\rm HeI} \Lambda_{\rm He} n_{\rm H}
        (f_{\rm He} - x_{\rm HeII})
        {\rm e}^{-h \nu_{{\rm HeI\,2^1}p\,2^1s}/k T_{\rm M}} 
        \right]}
       {H(z) (1+z) \left[ 1 + K_{\rm HeI}(\Lambda_{\rm He} 
        + \beta_{\rm HeI})n_{\rm H}
        (f_{\rm He} - x_{\rm HeII})
        {\rm e}^{-h \nu_{{\rm HeI\,2^1}p\,2^1s}/k T_{\rm M}} \right]} \; .
\end{equation}

The electron fraction is $x_e = x_p+x_{\rm HeII}$. The H 
${\rm Ly}\alpha$ wavelength is $\lambda_{{\rm H}\,2p} = 121.5682~{\rm 
nm}$. The H $2s-1s$ frequency is $\nu_{{\rm H}\,2s}$, nearly equal
to $c/\lambda_{{\rm H}\,2p}$. The \mbox{He\,{\sc i}} $2^1p-1^1s$
wavelength is $\lambda_{{\rm HeI}\,2^1p} = 58.4334~{\rm nm}$. The
\mbox{He\,{\sc i}} $2^1s-1^1s$ frequency is $\nu_{{\rm HeI}\,2^1s} =
c/60.1404~{\rm nm}$, and $\nu_{{\rm HeI}\,2^1p\,2^1s}$ is given by
$\nu_{{\rm HeI}\,2^1p\,2^1s} = \nu_{{\rm HeI}\,2^1p} - \nu_{{\rm
HeI}\,2^1s}$. The H $2s-1s$ two-photon rate is $\Lambda_{\rm H} =
8.22458~{\rm s^{-1}}$, and the \mbox{He\,{\sc i}} $2^1s-1^1s$
two-photon rate is $\Lambda_{\rm He} = 51.3~{\rm s^{-1}}$.
$K_{\rm H} \equiv \lambda_{{\rm H}\,2p}^3/[8 \pi H(z)]$ is the
cosmological redshifting of H ${\rm Ly}\alpha$ photons, which is the
reciprocal of $\Lambda_\alpha$ in Eq.~\ref{eq:CrFactor}. The
cosmological redshifting of \mbox{He\,{\sc i}} $2^1p-1^1s$ photons is
$K_{\rm HeI} \equiv \lambda_{{\rm HeI}\,2^1p}^3/[8 \pi H(z)]$. The
radiation temperature is given by $T_{\rm R} = T_0(1+z)$, with $T_0$ 
being the present CMB temperature.

\subsection{Matter Temperature}

The major effects affecting the evolution of matter temperature are
Compton cooling and adiabatic cooling. Other cooling and heating terms
have effects at $10^{-3}$\% level in the ionization fraction
\citep{SSS00}, and so we also ignore those terms in our calculation.

Electrons and photons are coupled by Compton scattering. When
electrons and photons are nearly in thermal equilibrium, the rate of
energy transfer is given by \citep{peebles68,peebles93,SSS00}
\begin{equation}\label{eq:comptonEnergy}
  \frac{{\rm d}E_e}{{\rm d}t}=\frac{4 \sigma_{\rm T} U n_e k}{m_e c} 
                  (T_{\rm R} - T_{\rm M}) \; ,
\end{equation}
where $E_e$ is the energy density of electrons and $\sigma_{\rm T}$ is
the Compton scattering cross section. The matter temperature therefore obeys:
\begin{equation}\label{eq:comptonTemp}
  \frac{{\rm d}T_{\rm M}}{{\rm d}t}=\frac{8}{3}
                        \frac{\sigma_{\rm T} U}{m_e c}
                        \frac{n_e}{n_{\rm tot}} (T_{\rm R} - T_{\rm M}) \; ,
\end{equation}
where $n_{\rm tot}$ is the total number density of particles.

We next consider the adiabatic cooling term. We treat the matter as an
ideal gas with an adiabatic index $\gamma = 5/3$. Recalling that $T_{\rm M}
\propto \rho^{\gamma-1}$ and $\rho \propto (1+z)^3$ (ignoring the
pressure of baryonic matter in the Friedmann equation\footnote{By
Friedmann equations, $\dot{\rho} c^2 a + 3(\rho c^2 + p)\dot{a} =
0$. Since the baryon pressure is much smaller than the energy density 
$\rho c^2$, even though the pressure of baryon will be halved during the
recombination process, it is still a good approximation to ignore the
pressure.}), it can be proved easily that the adiabatic cooling rate
is described by
\begin{equation}\label{eq:adiabaticTemp}
  \frac{{\rm d}T_{\rm M}}{{\rm d}t}=-3(\gamma-1)H(t)T_{\rm M}=-2H(t)T_{\rm M} \; .
\end{equation}
Therefore the evolution of the matter temperature with respect to redshift
$z$ is
\begin{equation}\label{eq:matterT}
  \frac{{\rm d}T_{\rm M}}{{\rm d}z}
  = \frac{8 \sigma_{\rm T} U}{3 H(z) (1+z) m_e c} \,
    \frac{n_e}{n_e+n_{\rm H}+n_{\rm He}} \,
    (T_{\rm M} - T_{\rm R}) +\frac{2 T_{\rm M}}{(1+z)} \; .
\end{equation}

By solving Eq.~\ref{eq:HRate}, Eq.~\ref{eq:HeRate} and
Eq.~\ref{eq:matterT}, the ionization fractions and matter temperature
can be found as a function of $z$.

\section{Recombination softening}\label{sec:modification}
In the derivation of Eq.~\ref{eq:matterT}, we have assumed that the
baryonic matter is an ideal gas. It is a good approximation when the
gas is either neutral or fully ionized. However, a partially ionized
gas tends to recombine under compression, and so the EOS
should be softer than that of an ideal gas. Here we rewrite the
temperature equation and investigate the effect.

Again we consider the recombination of H and \mbox{He\,{\sc i}}
only. The specific internal energy is the sum of the translation,
excitation, and ionization energies of the particles H and
\mbox{He\,{\sc i}} per gram of materials:
\begin{equation}\label{eq:internalEnergy1}
  e = \frac{1}{n_{\rm H}(m_{\rm H}+f_{\rm He} m_{\rm He})} 
      \left[ \frac{3}{2}(n_{\rm H}+n_e+n_{\rm He}) k T_{\rm M}
       + \sum_i n_{i \, {\rm H}} \, \epsilon_{i \, {\rm H}}
       + n_p \, \epsilon_{\rm H}
       + \sum_i n_{i \, {\rm HeI}} \, \epsilon_{i \, {\rm HeI}} 
       + n_{\rm HeII} \, \epsilon_{\rm HeI}\right] \; .
\end{equation}
Here we have assumed that all \mbox{He\,{\sc iii}} have already
recombined to \mbox{He\,{\sc ii}}, and so the total number density of
particles is $n_{\rm H} + n_e + n_{\rm He}$. $\epsilon_{i \, {\rm H}}$
is the excitation energy of the $i^{\rm th}$ excited state of H,
$\epsilon_{\rm H}$ is the ionization energy of H, $\epsilon_{i \, {\rm
HeI}}$ is the excitation energy of the $i^{\rm th}$ excited state of
\mbox{He\,{\sc i}}, and $\epsilon_{\rm HeI}$ is the ionization energy
of \mbox{He\,{\sc i}}.

Recalling the definition of $f_{\rm He} = n_{\rm He}/n_{\rm H}$, we
rewrite Eq.~\ref{eq:internalEnergy1} as
\begin{eqnarray}\label{eq:internalEnergy2}
  e&=& \frac{3}{2} \, (1+x_e+f_{\rm He}) \, 
       \frac{k}{m_{\rm H} + f_{\rm He} m_{\rm He}} 
       \, T_{\rm M}
       + \bigg[ \frac{1-x_p}{m_{\rm H} + f_{\rm He} m_{\rm He}}
       \sum_i \frac{n_{i \, {\rm H}}}{(1-x_p)n_{\rm H}} \, 
       \epsilon_{i \, {\rm H}}
       + \frac{x_p \, \epsilon_{\rm H}}
       {m_{\rm H} + f_{\rm He} m_{\rm He}}    \nonumber \\
   & & +\frac{f_{\rm He}-x_{i \, {\rm HeI}}}
       {m_{\rm H} + f_{\rm He} m_{\rm He} }
       \sum_i \frac{n_{i \, {\rm HeI}}}
       {(f_{\rm He}-x_{i \, {\rm HeI}})n_{\rm H}} \, 
       \epsilon_{i \, {\rm HeI}}
       +\frac{x_{\rm HeII} \, \epsilon_{\rm HeI}}
       {m_{\rm H} + f_{\rm He} m_{\rm He} } \bigg] \; .
\end{eqnarray}
The excitation energy terms in Eq.~\ref{eq:internalEnergy2} are found
to be negligible\footnote{To be explained in the appendix.}, and so we have
\begin{eqnarray}\label{eq:internalEnergy3}
  {\rm d}e &=& \frac{3}{2} \, (1+x_e+f_{\rm He}) \, 
       \frac{k}{m_{\rm H} + f_{\rm He} m_{\rm He}} \, {\rm d}T_{\rm M}
       + \, \left( \frac{3}{2} \, 
       \frac{k}{m_{\rm H} + f_{\rm He} m_{\rm He}} T_{\rm M}
       + \frac{\epsilon_{\rm H}}
       {m_{\rm H} + f_{\rm He} m_{\rm He}} \right)
       {\rm d}x_p                                        \nonumber \\
     & &  + \, \left( \frac{3}{2} \, 
       \frac{k}{m_{\rm H} + f_{\rm He} m_{\rm He}} T_{\rm M}
       + \frac{\epsilon_{\rm HeI}}
       {m_{\rm H} + f_{\rm He} m_{\rm He} } \right) 
       {\rm d}x_{\rm HeII} \; .
\end{eqnarray}

On the other hand, we can also write for an adiabatic process
\begin{equation}\label{eq:1stLawAdiabatic}
  {\rm d}e=\left(\frac{p}{\rho^2}\right) {\rm d}\rho \; ,
\end{equation}
where $p$ is the pressure, given by the ideal gas law\footnote{Since the 
recombination time scale is long compared to that to establish thermal
equilibrium in the cosmic plasma, the composition of the baryon gas
remains nearly constant. Due to this fact and also the fact that kinetic 
energy of baryons during recombination is much smaller than the excitational
energy of hydrogen, the ideal gas law is valid. However, the adiabatic 
index is changed because of the changing degrees of freedom during 
recombination.} as
\begin{eqnarray}\label{eq:idealGasLaw}
 p  =(1 + f_{\rm He} + x_p + x_{\rm HeII}) \, \rho \, 
      \frac{k}{m_{\rm H} + f_{\rm He} m_{\rm He}} \, T_{\rm M} \; .
\end{eqnarray}

Now we may eliminate $p$ to obtain an equation with $\rho$, $T_{\rm M}$,
$x_p$, $x_{\rm HeII}$ only. Note that our method is similar to that in
section~14 of \citet{mihalas}, but there is an important
difference. In \citet{mihalas}, the Saha equation is used to
eliminate the ionization fraction. We certainly should not use the
Saha equation in this case, but instead we will keep the derivative of
the ionization fraction.
Putting Eq.~\ref{eq:idealGasLaw} into Eq.~\ref{eq:1stLawAdiabatic} to
eliminate $p$, we get
\begin{equation}
  {\rm d}e=\left[(1 + f_{\rm He} + x_p + x_{\rm HeII}) \, 
     \frac{k}{m_{\rm H} + f_{\rm He} m_{\rm He}} \, T_{\rm M} \right] 
     \frac{{\rm d}\rho}{\rho} \; ,
\end{equation}
which can then be combined with Eq.~\ref{eq:internalEnergy3} and 
$\rho \propto (1+z)^3$ to obtain the modified temperature equation
\begin{eqnarray}\label{eq:newMatterT}
  \frac{1+z}{T_{\rm M}} \frac{{\rm d}T_{\rm M}}{{\rm d}z}
     &=&
         \frac{8 \sigma_{\rm T} U}{3 H(z) m_e c} \,
         \frac{n_e}{n_e+n_{\rm H}+n_{\rm He}} \,
         \frac{T_{\rm M} - T_{\rm R}}{T_{\rm M}}
        \, + \, 2 \,
        - \left(1+\frac{2}{3} \, \frac{\epsilon_{\rm H}}{k T_{\rm M}} \right)
          \frac{1+z}{1+f_{\rm He}+x_e} \, \frac{{\rm d}x_p}{{\rm d}z} \,
                                             \nonumber \\
     & &   - \left(1+\frac{2}{3} \, 
          \frac{\epsilon_{\rm HeII}}{k T_{\rm M}} \right)
          \frac{1+z}{1+f_{\rm He}+x_e} \, \frac{{\rm d}x_{\rm HeII}}{{\rm d}z} \,
           \; .
\end{eqnarray}

Eq.~\ref{eq:newMatterT} should then be used instead of the original equation
Eq.~\ref{eq:matterT}. Hereafter we call the first term on the right to
be the Compton Scattering term, the second the original adiabatic cooling 
term, the third and last the modified hydrogen and helium terms respectively.

\section{Results}\label{sec:results}

Now we investigate the effects of recombination softening and reheating. 
The model that we consider is the $\Lambda$CDM model, with dark energy, 
CDM, baryon and radiation as the components in the universe. Our choice 
of the cosmological parameters bases on the Running Spectral Index 
Model of {\it WMAP} \citep{wmapResult,wmapPara}. In this model, the 
universe is flat, and so the total density parameter $\Omega_0$ is 1.0 
and $\Omega_{\rm K}$ is 0.0. The contribution of the dark energy 
$\Omega_\Lambda$ is 0.73, and the total matter density $\Omega_{\rm M}$ 
is 0.27, in which the CDM density $\Omega_{\rm C}$ is 0.226 and the baryon 
density $\Omega_{\rm B}$ is 0.044. The EOS index of dark energy is equal 
to $-0.78$ and is fixed. We use $h=0.71$, $T_0=2.725$~K and $Y_{\rm p}=0.24$.
The scalar spectral index $n_{\rm s}$ is 0.93, and the derivative of 
spectral index ${\rm d}n_{\rm s}/{\rm d}\ln k$ is $-0.031$. We also 
simplify the calculation by assuming no reionization, lensing effect, 
tensor perturbations and massive neutrino in both models. Besides, we 
use {\it COBE} normalization throughout our calculation.

We have modified the code \mbox{\sc RECFAST} in \mbox{\sc CMBFAST}~4.3
\citep{seljak96}. The recombination process of \mbox{He\,{\sc iii}} to 
\mbox{He\,{\sc ii}} is calculated using the Saha equation. We solved a 
set of three differential equations, namely, the ionization equations for 
H and \mbox{He\,{\sc i}} and the modified temperature equation for matter
Eq.~\ref{eq:newMatterT}. The equations are then integrated by using the 
integrator \mbox{\sc DVERK} \citep{SSS99}. As in the original 
\mbox{\sc RECFAST}, we turn off \mbox{He\,{\sc ii}} whenever its 
ionization fraction is small enough. We use a fudge factor 1 in our
calculation. However, we find that the conclusion that we make below is
unchanged if we use the fudge facter used in \mbox{\sc RECFAST}.

\begin{figure}
 \centering
 \includegraphics[height=84mm,angle=0]{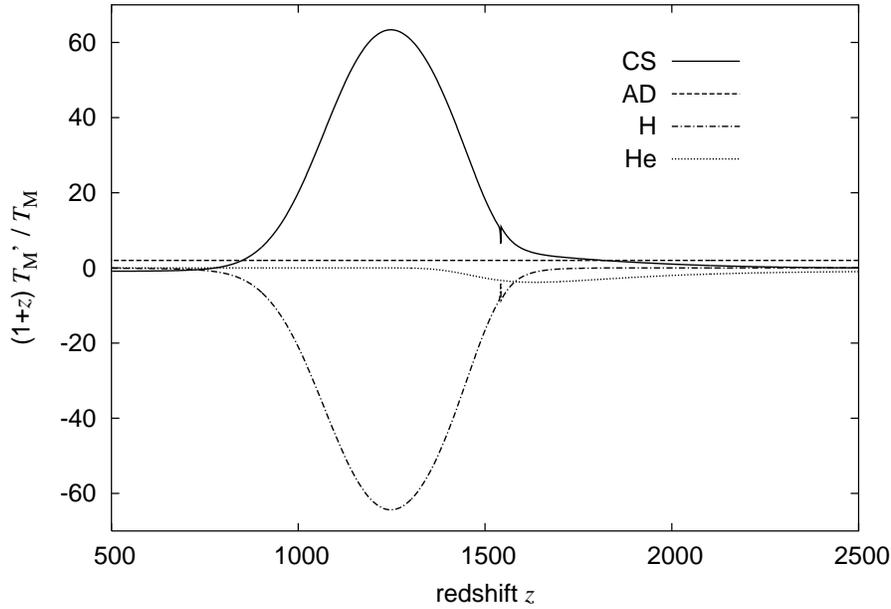}
 \caption{The $z$ dependence of the various terms of the modified temperature 
  equation Eq.~\ref{eq:newMatterT}. The solid, dashed, dot-dashed, and 
  dotted lines are the Compton Scattering (CS), adiabatic cooling (AD), 
  and the modified Hydrogen (H) and Helium (He) terms respectively.}
 \label{fig:fig1}
\end{figure}

Fig.~\ref{fig:fig1} shows the values of the various terms on the r.h.s.~of 
the modified temperature equation Eq.~\ref{eq:newMatterT}.  The 
recombination reheating tends to delay the recombination process. However, 
Compton scattering keeps the matter temperature close to the radiation
temperature, because the ionization fraction is still quite large. This 
explains the shapes of the curves in Fig.~\ref{fig:fig1}. This tight-coupling
effect in fact has reduced the effect of recombination softening and reheating.

\begin{figure}
 \centering
 \includegraphics[height=84mm,angle=0]{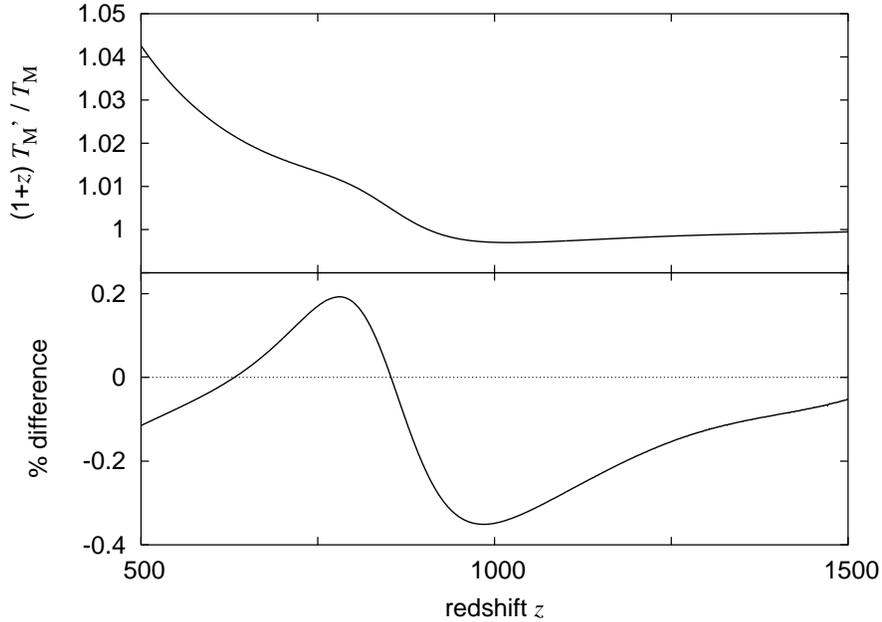}
 \caption{Sum of the terms on the r.h.s.~of the temperature equation 
  (upper panel) and the percentage difference with and without 
  recombination softening and reheating effects (lower panel). A fudge 
  factor of 1 has been used in the calculation.}
 \label{fig:fig2}
\end{figure}

The sum of all four terms in Eq.~\ref{eq:newMatterT} is plotted in the
upper panel of Fig.~\ref{fig:fig2}. At large $z$, the coefficient is 
equal to 1. This can be understood by knowing that matter and photons 
couple very well before recombination. Because the heat capacity of 
radiation is much larger than that of matter \citep{peebles93}, the 
matter temperature will follow that of photons. The value of the 
coefficient is thus equal to $3(\gamma-1)=3(4/3 - 1)=1$. The coefficient 
deviates from 1 as $z$ decreases. That is because the ionization 
fraction drops so fast that the Compton scattering can no longer force 
matter to follow the photon temperature. The percentage difference in 
$(1+z)\,{\rm d}(\ln T_{\rm M})/{\rm d}z$ with and without recombination 
softening and reheating effects is shown in the lower panel of 
Fig.~\ref{fig:fig2}. When the recombination starts, the sum of the terms 
is smaller than the original calculation; this means the temperature 
decreases more slowly after the modification.  

\begin{figure}
 \centering
 \includegraphics[height=84mm,angle=0]{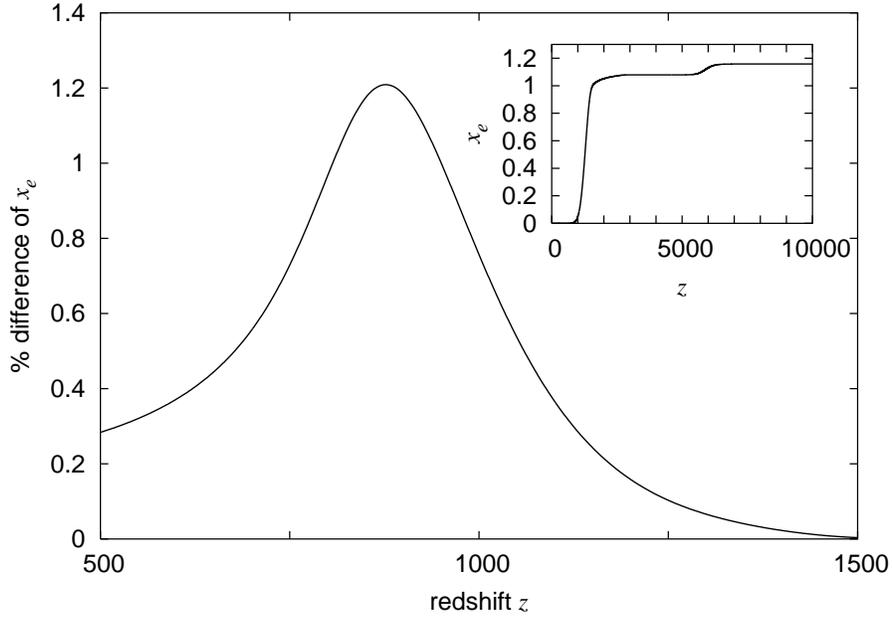}
 \caption{Change in the ionization fraction due to recombination softening
  and reheating. The subpanel shows the ionization history from between
  present and $z = 10000$.}
 \label{fig:fig3}
\end{figure}

The percentage difference in the ionization fraction with and without
recombination softening and reheating is plotted in Fig.~\ref{fig:fig3}. 
The original ionization history is shown as a subpanel. As the redshift 
decreases, the ionization fraction $x_e$ drops. The terms 
${\rm d}x_p/{\rm d}z$ and ${\rm d}x_{\rm HeII}/{\rm d}z$ in the new 
temperature equation are not zero for decreasing ionization fraction, 
and so the ionization fraction is also modified.

\begin{figure}
 \centering
 \includegraphics[height=84mm,angle=0]{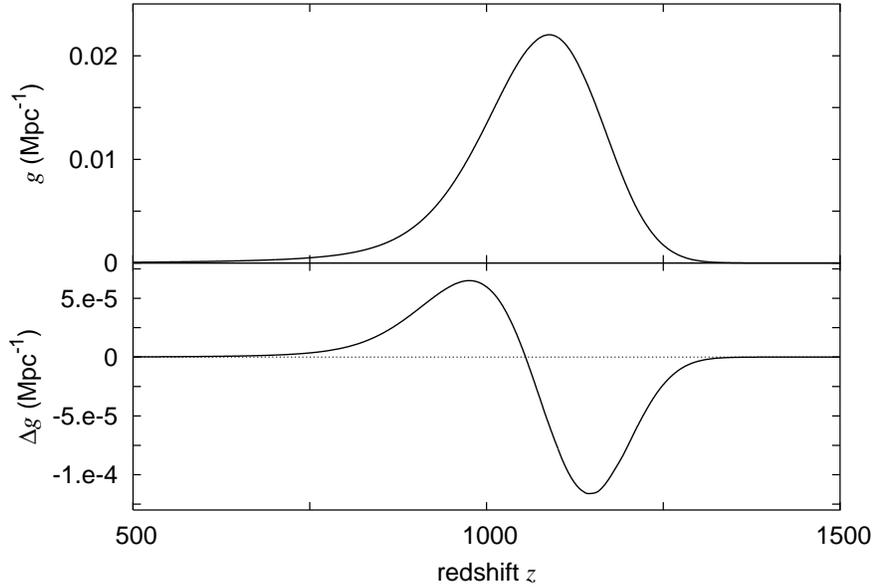}
 \caption{The visibility function $g$ and its change $\Delta g$ when 
  recombination softening and reheating are inluded. A fudge factor of 
  1 has been used in the calculation.}
 \label{fig:fig4}
\end{figure}

With the ionization fraction at different redshift $z$, we can then
calculate the visibility function $g(z)=\exp(-\tau){\rm d}\tau/{\rm d}\eta(z)$,
where $\tau$ is the optical depth, $\eta(z)$ is the conformal time. The 
peak of the visibility function defines the time of recombination, 
whereas its width defines the thickness of the last scattering 
surface. The visibility function and its changes are shown in 
Fig.~\ref{fig:fig4}, which clearly shows that the softening and 
reheating effect has delayed the recombination process.

We can understand the effect of the delay in recombination in terms 
of entropy. The specific entropy $s$ (entropy per unit mass) is 
defined by $T\,{\rm d}s = {\rm d}e+p\,{\rm d}(1/\rho)$.  For matter, the
change of specific entropy in the recombination process is
\begin{eqnarray}
  {\rm d}s &=& \frac{1}{T_{\rm M}} \left[{\rm d}e - 
                \left(\frac{p}{\rho^2}\right){\rm d}\rho \right]
                           \nonumber \\
     &=& \frac{k(1+x_p+x_{\rm HeII}+f_{\rm He})}
          {m_{\rm H}+f_{\rm He}m_{\rm He}}
         \bigg[\frac{3}{2} \, \frac{{\rm d}T_{\rm M}}{T_{\rm M}} 
         - 3 \, \frac{{\rm d}z}{1+z}
         + \left(\frac{3}{2}+\frac{\epsilon_{\rm H}}{k T_{\rm M}}
           \right)
           \frac{{\rm d}x_p}{1+x_p+x_{\rm HeII} +f_{\rm He}} \, \nonumber \\
     & &  + \left(\frac{3}{2}+\frac{\epsilon_{\rm He II}}{k T_{\rm M}} 
           \right) \, 
           \frac{{\rm d}x_{\rm He II}}{1+x_p+x_{\rm HeII}+f_{\rm He}}
           \bigg] \; .
\label{eq:change_specific_entropy}
\end{eqnarray}
A certain amount of entropy is released when the ions and electrons 
recombined. The terms ${\rm d}x_p$ and ${\rm d}x_{\rm He II}$ correspond to 
this effect. The entropy is then shared among the particles 
remaining in thermal contact, as reflected in the prefactor 
$1+x_p+x_{\rm He II}+f_{\rm He}$. As the gas recombines, the entropy 
is shared among less particles, thus the gas is heated\footnote{One
can also say that the cooling process is slowed down.} and the 
recombination process is delayed. 

\begin{figure}
 \centering
 \includegraphics[height=84mm,angle=0]{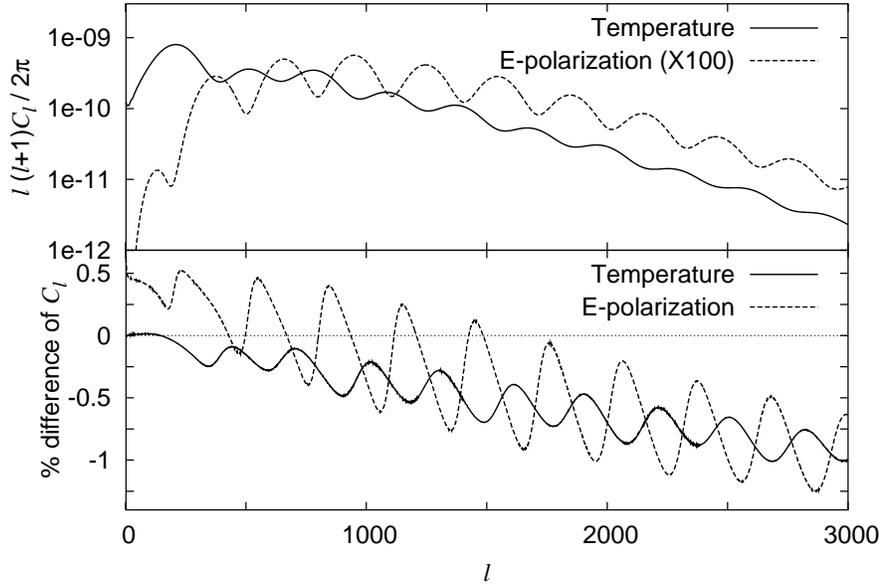}
 \caption{CMB temperature and E-polarization spectra (upper panel), and
  the percentage differences of the spectra $C_l$ (lower panel), 
  comparing results calculated with and without recombination softening 
  and reheating effects.}
 \label{fig:fig5}
\end{figure}

A delay of recombination changes the sound horizon at recombination, 
which directly affects the CMB power spectra. The new CMB power 
spectra and the percentage difference are shown in the lower panel of 
Fig.~\ref{fig:fig5}, with the original CMB spectra shown in the 
upper panel for comparison. Since the recombination is only delayed 
slightly, the change of peak positions is negligible. However, the 
change of amplitude of $C_l$ can be as high as 1\% to 2\%. In the 
results of {\it WMAP}, the temperature fluctuations $\Delta_T$ at the 
first peak, the next trough, and the second peak are $74.7\pm0.5~\mu$K,
$41.0\pm0.5~\mu$K, and $48.8\pm0.9~\mu$K respectively \citep{wmapResult}. 
They are equivalent to uncertainties of 1.3\%, 2.5\% and
3.5\% in $C_l$ at $l$ = 220.1, 411.7 and 546. Therefore, the error
of calculation due to ignoring the recombination softening and 
reheating effect is about 10 times smaller than error of {\it WMAP}. 
However, {\it WMAP} is an ongoing project and the observational uncertainties 
will be reduced. Also, the next generation of CMB observations, 
such as {\it PLANCK}, may achieve precision better than 1\% for 
$ l < 1000$ \citep{planckreport}, and so the recombination softening and 
reheating effects should be included in future analysis.

Besides experimental error in $C_l$, there is a statistical error 
known as cosmic variance. The theoretical limit of precision in the
determination of $C_l$ is $\sim$1\% at $l=100$, and $\sim$0.1\% at 
$l=1000$ \citep{araa}.  The effects of our modification are 
thus much larger than the cosmic variance for most $l$. Also, the 
effect of recombination softening and reheating increases as 
$\Omega_{\rm B}$ increases. This is expected since our modification of the 
temperature equation is solely for the baryons. However, even when 
$\Omega_{\rm B}$ is as large as 1, the change in $C_l$ is only about 
twice as in the $\Lambda$CDM model. Thus our conclusion is the same in 
a large range of $\Omega_{\rm B}$.

The change of the CDM transfer function $T(k)$ is of order 0.01\%. We
normalize the transfer function at large scale to unity, so by definition 
there is no change in the large scale (small $k$). As the
scale decreases, there is a region of $T(k)$ with signature of
fluctuations in photon-baryon fluid. The delay of recombination will
change the sound horizon and so will modify the fluctuations. At even
smaller scales (large $k$), the fluctuations are damped by Silk damping.
The damping length is about the geometric mean of the horizon size
and the mean free path \citep{peacock99}, and it is therefore larger in
our calculation. This explains the decrease in the transfer function that
we have found in our calculation. The new and old transfer functions 
approach a constant ratio because our calculation does not change 
the asymptotic behavior of the transfer function. The change in the 
transfer function is too small and can be safely neglected.

\section{Summary}\label{sec:summary}

We have presented all results using a fudge factor of 1 in the 
recombination calculation. We have also repeated the calculation using 
the fudge factor used in \mbox{\sc RECFAST}, and the conclusion is 
essentially the same. Whether the fudge factor itself 
should be changed when taking recombination softening and reheating 
into account awaits further calculation.

The relative difference of the ionization fraction $x_e$ calculated by
the modified \mbox{\sc RECFAST} with the new temperature equation and the
original \mbox{\sc RECFAST} increases as z decreases for $z<300$. However, 
the temperature equation breaks down for small redshift. As the universe 
continues to cool down, complications such as the formation of 
molecules, formation of structure at different scales, and reionization 
dominate the physics.

In summary, we have studied the effects of recombination softening and
reheating of the cosmic plasma on the ionization history, visibility 
function, the CMB spectra, and the CDM transfer function, 
using a generalized adiabatic index and standard cosmological parameters.  
With the standard cosmological parameters ($\Lambda$CDM model), the 
effects on the CMB spectra and ionization fraction are about 0.5\% 
at $l=1500$ and about 1\% at $l=3000$. These numbers are small compared 
to the current observational uncertainties but are not negligible. As
the experimental errors become smaller, this improvement 
of the calculation should be used for determining the cosmological
parameters. The effect on the CDM transfer function is much 
smaller, and so this modification is not important in the calculation 
of matter power spectrum.

The work described in this paper was substantially supported by a grant
from the Research Grants Council of the Hong Kong Special Administrative
Region, China (Project No. 400803), and a Direct Grant (Project Code:
2060248) from the Chinese University of Hong Kong.


\section{Appendix}

\subsection{Comparsion of the energy terms}

For the density and temperature of the baryon gas at recombination,
the major effect of line broadening is thermal broadening. The gas is
nearly in thermal equilibrium and so follows Maxwell's distribution. 
The broadening is therefore about the width of the gaussian function. 
See \citet{kenneth80} for a more detailed discussion of the effect.

The thermal broadening can be calculated by $\nu (2 k T_{\rm M}
/m_{\rm H} c^2)^{1/2}$, where $\nu$ is the photon frequency. 
\citet{SSS00} has shown that at recombination the
energy separation between the $n=300$ level and the continuum of
hydrogen is nearly equal to the energy width of the thermal
broadening. The electrons in level higher than 300 will then be
ionized thermally. Similarly we should include only 300 levels in our
calculation of the excitation energy of hydrogen. For helium, we should
follow \citet{SSS00} to use 200 levels. Without this truncation of
states, the sum in the excitation terms will diverge unphysically.

After the \mbox{He\,{\sc ii}} recombination, the specific internal 
energy of baryonic matter is
\begin{equation}
  e= \frac{1}{n_{\rm H}\left(m_{\rm H}
        +f_{\rm He}m_{\rm He}\right)}
     \bigg[ \frac{3}{2} n_{\rm H}
        (1+x_p+x_{\rm He II}+f_{\rm He}) k T_{\rm M}
        + \sum_i n_{i {\rm H}} \epsilon_{i {\rm H}}
        + n_p \epsilon_{\rm H}
        + \sum_i n_{i {\rm He I}} \epsilon_{i {\rm He I}}
        + n_{\rm He II} \epsilon_{\rm He I} \bigg] \; ,
\label{eq:internal_energy}
\end{equation}
where the first term is the thermal energy of all particles, the
second and the third terms are the excitation energy and ionization
energy of hydrogen atom respectively, and the fourth and fifth terms 
are the excitation energy and ionization energy of \mbox{He\,{\sc i}}.

We are able to calculate the two ionization energy terms. We now
simplify the hydrogen excitation energy term. When H starts to
recombine, He has nearly recombined completely. The number density of H
atoms with the electron in the $i^{\rm th}$ state is given by
\citep{mihalas}
\begin{equation}
  n_{i {\rm H}}=2 C i^2 n_e n_p T_{\rm M}^{-3/2} 
      {\rm e}^{\epsilon_{\rm H}/i^2 k T_{\rm M}} \; ,
\label{eq:ith_state}
\end{equation}
with $C=(1/2)(h^2 / 2 \pi m_e k)^{3/2}$.

The excitation energy in the $i^{\rm th}$ state of hydrogen is given
by $\epsilon_{i{\rm H}}=(1-1/i^2)\epsilon_{\rm H}$, and so the
excitation energy term of hydrogen is
\begin{eqnarray}
  \frac{1}{n_{\rm H}\left(m_{\rm H}
        +f_{\rm He}m_{\rm He}\right)}
  \sum_i n_{i {\rm H}} \epsilon_{i {\rm H}}
    &=& \frac{1-x_p}{m_{\rm H}+f_{\rm He}m_{\rm He}}
         \sum_i\left(\frac{n_{i {\rm H}}}{n_{\rm H}-n_p}\right)
         \epsilon_{i{\rm H}}                         \nonumber \\
    &=& \frac{1}{m_{\rm H}+f_{\rm He}m_{\rm He}} \,
         \frac{2 C \rho_{\rm cr}\Omega_{\rm B}(1-Y_{\rm p})\epsilon_{\rm H}}
         {m_{\rm H}}
         x_p^2 (1+z)^3 T_{\rm M}^{-3/2}
         \sum_i(i^2-1) {\rm e}^{\epsilon_{\rm H}/i^2 k T_{\rm M}} \; ,
\end{eqnarray}
where $\rho_{\rm cr}$ is the critical density.

The excitation energy for \mbox{He\,{\sc i}} is much more difficult to
calculate. The helium atom is not hydrogenic. There is no close form
for the energy states. One has to use methods like perturbation method
or variational method. To calculate the energy states with high
accuracy, one can refer to \citet{drake96}.

However for our purpose of checking the validity of neglecting the
excitation terms, we do not have to calculate the correct energy
states. Instead we find an upper limit of the states. We calculate the
\mbox{He\,{\sc i}} excitation energy with similar steps as the
hydrogen. Then we construct a hydrogenic atom with a helium nucleus 
and one electron. Because we have neglected the repulsive force of
another electron, the calculated energy in a certain state will be 
larger than the energy in the same state of the real helium atom. 
Therefore we can find a upper limit for the \mbox{He\,{\sc i}} 
excitation energy.

\begin{figure}
 \centering
 \includegraphics[height=84mm,angle=0]{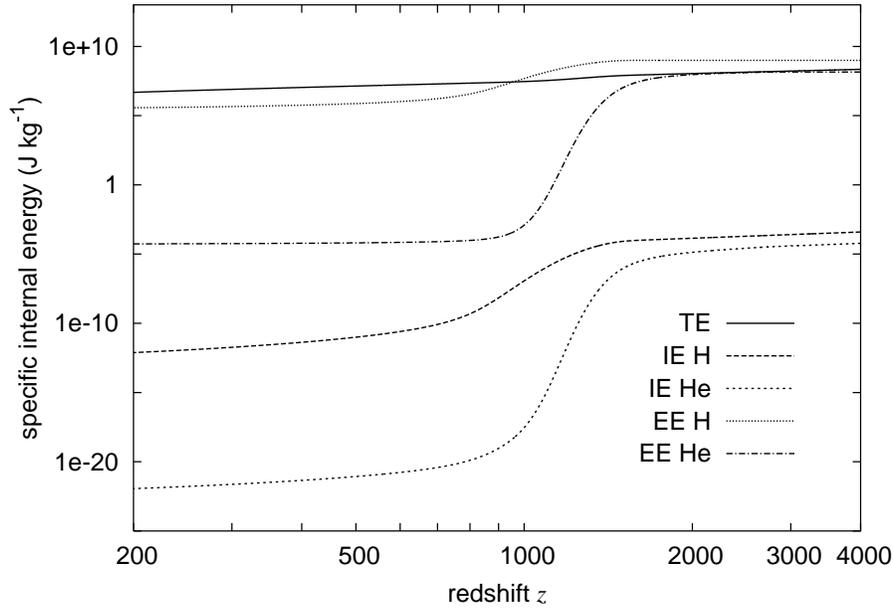}
 \caption{Different terms in the specific internal energy. The lines 
  represent contribution from thermal energy of all particles (TE), 
  ionization energy of hydrogen (IE H), ionization energy of helium~I 
  (IE He), excitation energy of hydrogen (EE H), and the upper limit of 
  the excitation energy of helium~I (EE He).}
 \label{fig:fig6}
\end{figure}

Fig.~\ref{fig:fig6} shows the five terms of the
specific internal energy in Eq.~\ref{eq:internal_energy}. The two
excitation energy terms are much smaller than the thermal energy of
all particles and the ionization energy of hydrogen, and so it is
appropriate to neglect the excitation energy terms. The \mbox{He\,{\sc i}}
ionization energy term is kept because it is important before and 
during the \mbox{He\,{\sc i}} recombination.


\bsp

\label{lastpage}


\begin{thebibliography}{}
\bibitem[\protect\citeauthoryear{Bennett et al.}{2003}]{wmapResult}
 Bennett C. L., et al., 2003, ApJS, 148, 1
\bibitem[\protect\citeauthoryear{Spergel et al.}{2003}]{wmapPara}
 Spergel D. N., et al., 2003, ApJS, 148, 175
\bibitem[\protect\citeauthoryear{Bersanelli et al.}%
 {1996}]{planckreport} Bersanelli M., et al., 
 Report on the Phase A Study of {\it COBRAS}/{\it SAMBA}
 ({\it PLANCK} project was originally named {\it COBRAS}/{\it SAMBA})
\bibitem[\protect\citeauthoryear{Chandrasekhar}%
 {1939}]{chandrasekhar39} Chandrasekhar S., 1939, An Introduction 
 to the Study of Stellar Structure, Univ. of Chicago Press, 
 Chicago
\bibitem[\protect\citeauthoryear{Drake}{1996}]{drake96}
 Drake G. W. F., 1996, Atomic, Molecular, \& Optical
 Physics Handbook, AIP Press, Woodbury, NY, chapter 11
\bibitem[\protect\citeauthoryear{Durrer}%
 {2001}]{durrer} Durrer R., 2001, J. Phys. Stud., 5, 177
\bibitem[\protect\citeauthoryear{Hannestad}{2001}]{steen}
 Hannestad S., 2001, New Astronomy, 6, 17
\bibitem[\protect\citeauthoryear{Hu \& Dodelson}%
 {2002}]{araa} Hu W., Dodelson S., 2002, ARA\&A, 40, 171
\bibitem[\protect\citeauthoryear{Hu et al.}{1995}]{HSSW95}
 Hu W., Scott D., Sugiyama N., White M., 1995, ApJ, 52, 5498
\bibitem[\protect\citeauthoryear{Lang}{1980}]{kenneth80}
 Lang K. R., 1980, Astrophysical Formulae, 2nd edition,
 Springer-Verlag, Berlin, section 2.18
\bibitem[\protect\citeauthoryear{Ma \& Bertschinger}%
 {1995}]{ma95}Ma C.-P., Bertschinger E., 1995, ApJ, 455, 7
\bibitem[\protect\citeauthoryear{Mihalas \& Mihalas}{1999}]{mihalas}
 Mihalas D., Mihalas B. W., 1999, Foundations of 
 Radiation Hydrodynamics, Dover, Mineola, NY
\bibitem[\protect\citeauthoryear{Peacock}{1999}]{peacock99}
 Peacock J. A., 1999, Cosmological Physics, Cambridge Univ. Press,
 Cambridge, UK
\bibitem[\protect\citeauthoryear{Peebles}{1968}]{peebles68}
 Peebles P. J. E., 1968, ApJ, 153, 1
\bibitem[\protect\citeauthoryear{Peebles}{1993}]{peebles93}
 Peebles P. J. E., 1993, Principles of Physical Cosmology, 
 Princeton Univ. Press, Princeton, NJ
\bibitem[\protect\citeauthoryear{Seager et al.}%
 {1999}]{SSS99} Seager S., Sasselov D. D., Scott D., 1999, ApJ,
 523, L1
\bibitem[\protect\citeauthoryear{Seager et al.}%
 {2000}]{SSS00} Seager S., Sasselov D. D., Scott D., 2000, 
 ApJS, 128, 407
\bibitem[\protect\citeauthoryear{Seljak \& Zaldarriaga}%
 {1996}]{seljak96} Seljak U., Matias Z., 1996, ApJ, 469, 437
\end{thebibliography}
\end{document}